\documentclass[aps,prb,preprint,floatfix]{revtex4-1}
\usepackage{times}
\usepackage{color}
\usepackage[dvips]{graphicx}
\usepackage{amsmath}
\usepackage{amsfonts}
\usepackage{amssymb}
\begin{document}

\title{Morphological instability of core-shell metallic nanoparticles}

\author{Davide Bochicchio and Riccardo Ferrando\footnote{Corresponding author, e-mail: ferrando@fisica.unige.it}}
\affiliation{Dipartimento di Fisica and IMEM/CNR, Via Dodecaneso 33, Genova, I16146, Italy}

\begin{abstract}
Bimetallic nanoparticles (often known as nanoalloys) with core-shell arrangement are  of special interest in several applications, such as in optics, catalysis, magnetism and biomedicine. Despite wide interest in applications, the physical factors stabilizing the structures of these nanoparticles are still unclear to a great extent, especially for what concerns the relationship between geometric structure and chemical ordering pattern. Here global-optimization searches are performed in order to single out the most stable chemical ordering patterns corresponding to the most important geometric structures, for a series of weakly miscible systems, including AgCu, AgNi, AgCo and AuCo. The calculations show that (i) the 
overall geometric structure of the nanoalloy and the shape and placement of its inner core are strictly correlated; (ii) centered cores can be obtained in icosahedral nanoparticles but not in crystalline or decahedral ones, in which asymmetric quasi-Janus morphologies form;  (iii) in icosahedral nanoparticles, when the core exceeds a critical size, a new type of morphological instability develops, making the core asymmetric and extending it towards the nanoparticle surface; (iv) multi-center patterns can be obtained in polyicosahedral nanoalloys. Analogies and differences between the instability of the core in icosahedral nanoalloys and the Stranski-Krastanov instability occurring in thin-film growth are discussed. All these issues are crucial for designing strategies to achieve effective coatings of the cores.
\end{abstract}

\maketitle

\section{Introduction}

Nanoalloys are  bi- or multi-component metallic particles  in the size range between 1-100 nm \cite{Ferrando08cr}. These nanoparticles 
are of great interest  for  both basic science and technological applications to magnetism, catalysis and optics. This wealth of possible applications stem from the very high degree of tunability of the physical and chemical properties of these systems, which is a direct consequence of the great variety of morphologies that nanoalloys can assume. 
Nanoalloys can present a variety of geometric shapes, ranging from crystalline structures, i.e. fragments of bulk crystals, to non-crystalline structures, the most common being icosahedra, decahedra and polyicosahedra \cite{Rossi04prl}. But variety is not limited to the geometric shape, since the atomic arrangement in nanoalloys can form qualitatively different chemical ordering patterns. 
Chemical ordering patterns can be intermixed (randomly or phase ordered), core-shell and multishell. Also phase separated patterns have been found, such as the so-called as Janus particles \cite{Parsina,Langlois08jnr}, and the ball-and-cup configurations \cite{Nunez2010jpcc,Bochicchio2012epjd,Langlois2012nanoscale}. 
This variety of geometries and patterns reflects the complexity of nanoalloy energy landscapes \cite{Walesbook}, which renders the theoretical prediction of most stable structures quite challenging. The determination of the most stable structures is the starting point for calculating the properties of nanoalloys, that in general depend both on geometry and on chemical ordering. The interplay of geometry and chemical ordering is far from trivial, and it is the subject of active investigation.

Silver- and gold-containing nanoalloys are presently attracting notable interest. In particular, the cases of poorly miscible pairs, such as
AgCu, AgCo, AgNi, and AuCo, have been investigated, both by experiments and simulations, by several groups that have explored a variety of properties and potential applications \cite{Rossi04prl,Baletto03prl,Baletto02prb,Baletto04ss,Janssens2005epjd,Rapallo05jcp,Mariscal05jcp,Berthier06jcp,Ortigoza08,Delogu2008prb,Calvo08prb,Lequien08prb,Lequien08fd,Langlois08fd,Delfour09prl,Delahoz09,Johnson09,Bochicchio2010nl,Kilimis2010,Nunez2010jpcc,Tang2011jcp,Weissker2011prb,Molayem2011jpccaAgNi,Molayem2011jpccaAgCu,Bochicchio2012epjd,Langlois2012nanoscale,Yildirim2012jpcc,Cune2012cpc,Shin2012ct,Xiao2012cpb,Rapallo2012jpcc}. 
AgCo, AgNi, and AuCo nanoalloys are studied for their interesting magneto-optical properties which stem from the fact that they are composed by a ferro- (Co or Ni) and a non-ferromagnetic metal (Ag or Au), the latter presenting a sharp surface plasmon resonance (SPR). The frequency of the SPR in these nanoalloys depends on composition and ordering pattern, so that it can be tuned to a good extent \cite{Gaudry03}. Applications to catalysis have been also proposed. For example, AgCo nanoalloys have been employed in the oxygen reduction in alkaline media \cite{Lima06}, and AgCu clusters are proposed as more efficient catalysts than the corresponding pure clusters for oxygen reduction reactions \cite{Shin2012ct}. AgCu nanoalloys find application in the fabrication of Pb-free solder interconnects, that can melt at low temperatures \cite{Kim10apl}. Core-shell Au-Co nanoparticles have been proposed  also for biomedical applications, such as thermal ablation therapies and drug delivery, for which the magnetic Co 
core 
is coated by a shell of Au, which is a biocompatible material that can be easily functionalized \cite{Bao2007jpcc,YangLu2010advmat,Chen2003jap}.

All these systems present some common features. Besides being poorly miscible in bulk phases \cite{Hultgren}, these systems have the element of larger atomic radius (Ag or Au) presenting smaller cohesive and surface energies than the other element. These features are all in favour of Ag or Au surface segregation in these nanoparticles, a behaviour that has been confirmed, both experimentally and computationally, by several studies \cite{Rossi04prl,Baletto03prl,Baletto02prb,Baletto04ss,Rapallo05jcp,Mariscal05jcp,Berthier06jcp,Ortigoza08,Delogu2008prb,Lequien08prb,Lequien08fd,Langlois08fd,Delfour09prl,Delahoz09,Johnson09,Bochicchio2010nl,Nunez2010jpcc,Molayem2011jpccaAgNi,Molayem2011jpccaAgCu,Bochicchio2012epjd,Langlois2012nanoscale,Yildirim2012jpcc,Cune2012cpc,Rapallo2012jpcc}. However, the fact that the cluster surface is expected to contain mostly Ag or Au does not determine completely chemical ordering. As we will see in the following, the latter is strongly influenced by a complex interplay with the 
nanoparticle geometrical shape.   

Generally speaking, even though several interesting results have been produced, the relation between geometry and chemical ordering in determining the most stable nanoparticle configuration is still to be understood in all these nanoalloys. In particular, it is still unclear how the geometric shapes influence the preferential chemical ordering.  In this paper we propose a systematic computational investigation of this topic, for AgCu, AgNi, AgCo, and AuCo nanoalloys. We consider Ag- or Au-rich compositions, that, as we will see, present the most complex and interesting interplay between shapes and patterns. From the computational point of view, most optimization studies about these nanoalloys have treated so far small cluster sizes, below 100 atoms, whose experimental observation at the electron microscope is quite challenging. There are very few global optimization studies for sizes above 100 atoms for these systems \cite{Bochicchio2010nl,Bochicchio2012epjd}. These studies are however 
focused on a specific structural motif (the chiral icosahedron) that is relevant for intermediate and Ag- or Au-poor compositions.  Here we consider clusters of sizes up to more than 10$^3$ atoms (diameters up to 3-4 nm), that can be more easily observed in electron microscopy experiments. 
This is especially interesting because techniques such as STEM with Z-contrast imaging \cite{Voyles2002nature} and energy-filtered TEM \cite{Cazayous2006prb}  allow to reveal the internal structures of binary nanoparticles. 
We show that shape and placement of the inner core is clearly correlated with the overall nanoparticle shape, centered cores being possible in icosahedra but not in decahedra and fcc nanocrystals. In icosahedra, the core is centered until it reaches a critical size at which a morphological instability develops. 
Finally, we show that multicenter cores are found in  polyicosahedra. 

\section{Model and Methods}
 
 From now on, the label S will refer to shell atoms (i.e. Ag or Au) and the label C to core atoms (Cu, Ni or Co). 
 
The nanoparticles are described within an atomistic model developed within the second-moment approximation to the tight-binding model (SMATB potential), known also as Gupta or RGL potential \cite{Cyrotlackmann71,Rosato89,Gupta81}.
In this model, the potential energy of the system depends on the relative distances between atoms $r_{ij}$ and it is  written as the sum of single-atom contributions $E_i$: 
\begin{equation}
 E_i= \sum_{j,r_{ij} \leq r_c} A_{ij} \mathrm{e}^{-p_{ij} \left(\frac{r_{ij}}{r_{0ij}} - 1 \right )} -
 \sqrt{\sum_{j,r_{ij} \leq r_c} \xi^2_{ij} \mathrm{e}^{-2 q_{ij} \left(\frac{r_{ij}}{r_{0ij}} - 1 \right )}}.
\end{equation} 
The parameters $p_{ij},q_{ij},A_{ij},\xi_{ij},r_{0ij}$ depend indeed on the atomic species of the pair only. Therefore $p_{ij}$ can be either  $p_{SS}$ or $p_{CC}$ or $p_{SC}$, and the same holds for all other parameters. As for $r_{0ij}$, $r_{0CC}$ and $r_{0SS}$ are the equilibrium distances in the bulk pure crystals at zero temperature, and $r_{0SC}=(r_{0CC}+r_{0SS})/2$. $r_c$ is an appropriate cutoff distance so that the sum has contributions from all neighbors of atom $i$ that are within $r_c$. In this work we choose to put $r_c$ equal to the second-neighbor distance in the respective bulk solids for SS and CC pairs, whereas for SC interactions the cutoff distance is the arithmetic average. The potential is then linked to zero at the third-neighbor distance by a polynomial function in such a way that the resulting function is continuous with continuous derivatives. The parameters of the potential can be found in \cite{Baletto02prb,Baletto03prl} for AgCu and AgNi, 
in \cite{Rossi09jctn} for AgCo and in \cite{Rapallo2012jpcc} for AuCo. The interaction potential has been favorably checked against density-functional calculations \cite{Bochicchio2010nl,Rapallo2012jpcc}.  

We note that all systems present a considerable lattice mismatch between the metals. If we measure the mismatch by the quantity $(r_{0SS}-r_{0CC})/r_{0SS}$ we obtain 0.118 for AgCu, 0.138 for AgNi, 0.135 for AgCo and 0.131 for AuCo.
 
In the following we will calculate the local pressure. For atom $i$, the local pressure $P_i$ is given by 
\begin{equation}
  P_i = - \frac{dE_i}{dV_i}, 
  \label{pressure}
\end{equation}
where $E_i$ is the atomic energy and $V_i$ is the atomic volume. $P_i$ is expressed as a function of the atomic coordinates as
\begin{equation}
  P_i = - \frac{1}{3 V_i}  \sum_{k \neq i} r_{ik} \left [ \frac{d \varphi_{ik}}{d r_{ik}} 
- \psi_{ik} \frac{d \psi_{ik}}{d r_{ik}} D(i) \right]
\end{equation}
where $V_i$ is the atomic volume in the bulk solid and
\begin{equation}
D(i) = \frac{1}{{\sqrt{\sum_{j \neq i} \psi_{ij}^2(r_{ij}) }}}.
\end{equation}
The functions $\varphi_{ij}$ and $\psi_{ij}$ are given by
\begin{eqnarray}
 \varphi_{ij}(r_{ij}) & = &  A_{ij} \mathrm{e}^{-p_{ij} \left(\frac{r_{ij}}{r_{0ij}} - 1 \right )} \nonumber \\
 \psi_{ij}(r_{ij}) & = &  \xi_{ij} \mathrm{e}^{-q_{ij} \left(\frac{r_{ij}}{r_{0ij}} - 1 \right )}
 \label{pressurefin}
\end{eqnarray}
if $r_{ij}$ is within the second-neighbor distance, by a polynomial link between the second- and third-neighbor distance and zero elsewhere.

The lowest energy structures are searched for by a basin-hopping procedure \cite{Wales1997} with exchange moves only. In an exchange move, the positions of two atoms of different species are swapped. The move is followed by local relaxation to reach the closest local minimum. Both random and tailored exchanges are used \cite{Bochicchio2010nl}. In tailored exchanges, atoms are chosen with different probabilities according to their local environment. Usually tailored exchanges allow finding chemical ordering arrangements with lower energy than random exchanges. Random exchanges become not efficient at all when dealing with the nanoparticle sizes considered in this work. It has been checked however that at smaller size, both random and tailored exchanges lead to the same results, the latter being faster in reaching the lowest-energy structures. For each size and composition, at least five independent simulations of 10$^{5}$ steps each have been made.

\section{Results and discussion}

\begin{figure} [ht]
\includegraphics[width=11cm]{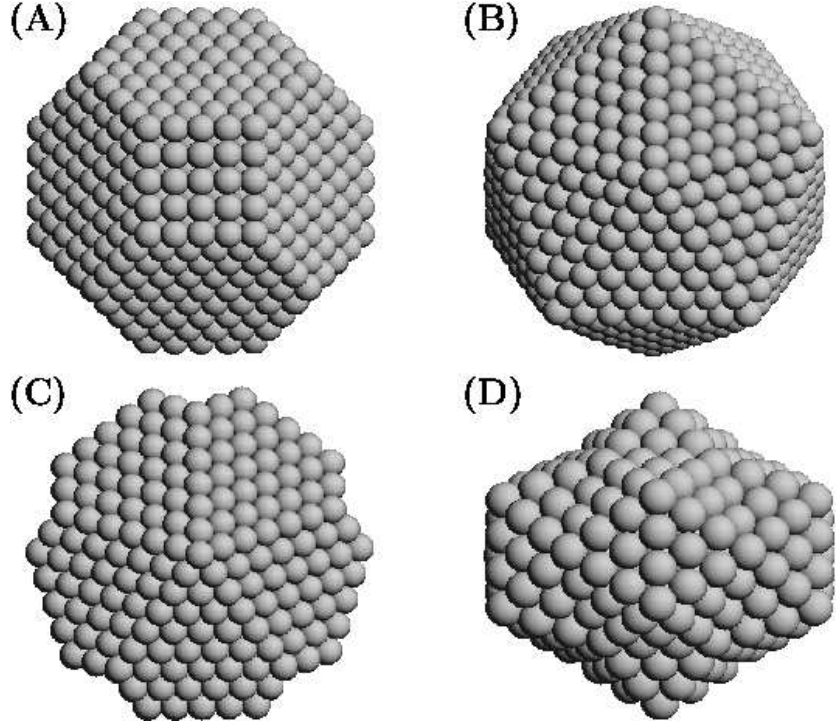}
 \caption{\label{fig_geom} (A) Fcc truncated octahedron of 1289 atoms. (B) Mackay icosahedron of 1415 atoms. (C) Marks decahedron of 1228 atoms. (D) Polyicosahedron of 419 atoms. It is made of five interpenetrating icosahedra of 147 atoms.}
\end{figure}

The main geometrical motifs that have been singled out in AgCu, AgCo, AgNi and AuCo clusters are fcc nanocrystals, icosahedra, decahedra, and polyicosahedra \cite{Ferrando08cr,Rossi04prl,Rapallo05jcp,Bochicchio2010nl,Molayem2011jpccaAgCu,Molayem2011jpccaAgNi}. Representative clusters of these geometrical motifs are shown in Fig. \ref{fig_geom}. In the following, we consider each geometrical motif and we look for the most energetically stable chemical ordering within that motif, for a series of selected compositions. Our global optimization of the chemical ordering is made by means of the algorithms described in the Methods section, within the atomistic potential model whose form and parameters are given there. 

\subsection{Fcc structures}

This motif has been studied in a series of computational works \cite{Baletto03prl,Calvo08prb,Parsina} showing that in AgCu, AgNi and AgCo the cores assume preferentially off-center positions in clusters in the size range of 2 nm. This has been recently confirmed experimentally for AgCu  \cite{Langlois2012nanoscale} in the case of larger nanoparticles, with diameters in the range 5 -10 nm. As we show in the following, our calculations reproduce this behavior for sizes close to the experimental ones, and for all systems including AuCo. We consider specifically a truncated octahedron of 1289 (diameter $\sim$4 nm) atoms and change the proportion of C atoms, explicitly considering the cases of 1, 2 , 5, 20 and 200 C atoms. 

\begin{figure} [ht]
\includegraphics[width=11cm]{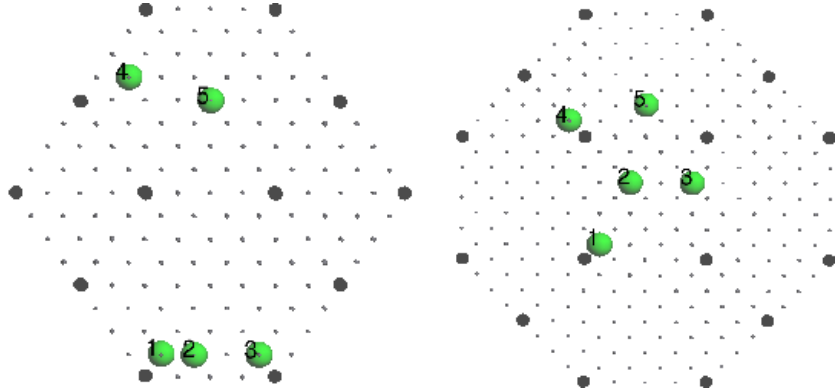}
 \caption{\label{fig_to1289_imp} Single-impurity sites in a truncated octahedron  of 1289 atoms considered in Table \ref{tablefcc}. The vertices of the truncated octahedron are represented by midsize spheres, whereas other atoms than the impurity are shown as small spheres.}
\end{figure}

For a single C atoms, the most favorable positions are susburface sites in all cases (see the data reported in Table \ref{tablefcc}, referring to the impurity sites of Fig. \ref{fig_to1289_imp}). This confirms what was already found for AgCu and AgNi fcc clusters by means of atomistic calculations \cite{Baletto03prl} and for AgCu clusters by DFT calculations \cite{Langlois2012nanoscale}. The most favorable position is always the subsurface site below a cluster vertex, with energy gains with respect to central inner sites that vary from a minimum of 0.057 eV for AuCo to a maximum of 0.170 eV for AgCo. Surface sites are disfavored compared to subsurface and inner sites, especially in AuCo. The energetic cost of bringing an impurity atom from an inner site to the center of a (100) facet varies from 0.05 eV in AgCo, to 0.09 eV in AgCu, to about 0.15 eV in AgNi and to more than 0.3 eV in AuCo. Therefore in AgCo the energetic cost for placing atoms in the cluster surface is especially small. This has some 
important consequences that we will analyze in the following.  

\begin{table}  [ht]
 \begin{tabular}{l|c|c|c|c|c|c|c|c}
         &  AgCu & AgCu & AgNi &  AgNi &  AgCo &  AgCo & AuCo & AuCo  \\
position &  $E$  & $P$  & $E$  &  $P$  &   $E$ &   $P$ &  $E$ &  $P$  \\
  \hline
surface, center of a (100) face                  &  0.087  &  -6.59 &  0.146  &  -6.82  &  0.050 &  -5.27  &  0.324  & -12.41 \\
surface, center of a (111) face                  &  0.083  &  -7.48 &  0.141  &  -9.17  &  0.050 &  -7.51  &  0.285  & -12.33 \\
inside the cluster, central site                 &  0.000  & -10.05 &  0.000  & -13.76  &  0.000 & -11.22  &  0.000  & -15.40 \\
1 subsurface, below a vertex site                & -0.068  &  -6.46 & -0.138  & -10.21  & -0.170 &  -6.88  & -0.057  &  -9.51 \\
2 subsurface, below the center of a (100) face   & -0.029  &  -8.29 & -0.062  & -11.98  & -0.081 &  -9.18  & -0.001  & -12.81 \\
3 subsurface, below the edge of a (100) face     & -0.049  &  -7.38 & -0.101  & -11.10  & -0.126 &  -7.97  & -0.030  & -11.19 \\
4 subsurface, below the edge of a (111) face     & -0.042  &  -8.12 & -0.089  & -11.85  & -0.100 &  -9.04  & -0.050  & -12.54 \\
5 subsurface, below the center of a (111) face   & -0.026  &  -9.01 & -0.051  & -12.78  & -0.056 & -10.23  & -0.033  & -13.78 \\
\hline  
\end{tabular}
\caption{\label{tablefcc} Energy $E$ and local pressure $P$ of a single impurity in some different positions in a truncated octahedron of 1289 atoms, after local relaxation. The energy in the central position inside the cluster is set to zero in all systems. Subsurface positions are numbered as in Fig. \ref{fig_to1289_imp}. Energies are given in eV, local pressures in GPa.}
\end{table}

\begin{figure} [ht]
\includegraphics[width=11cm]{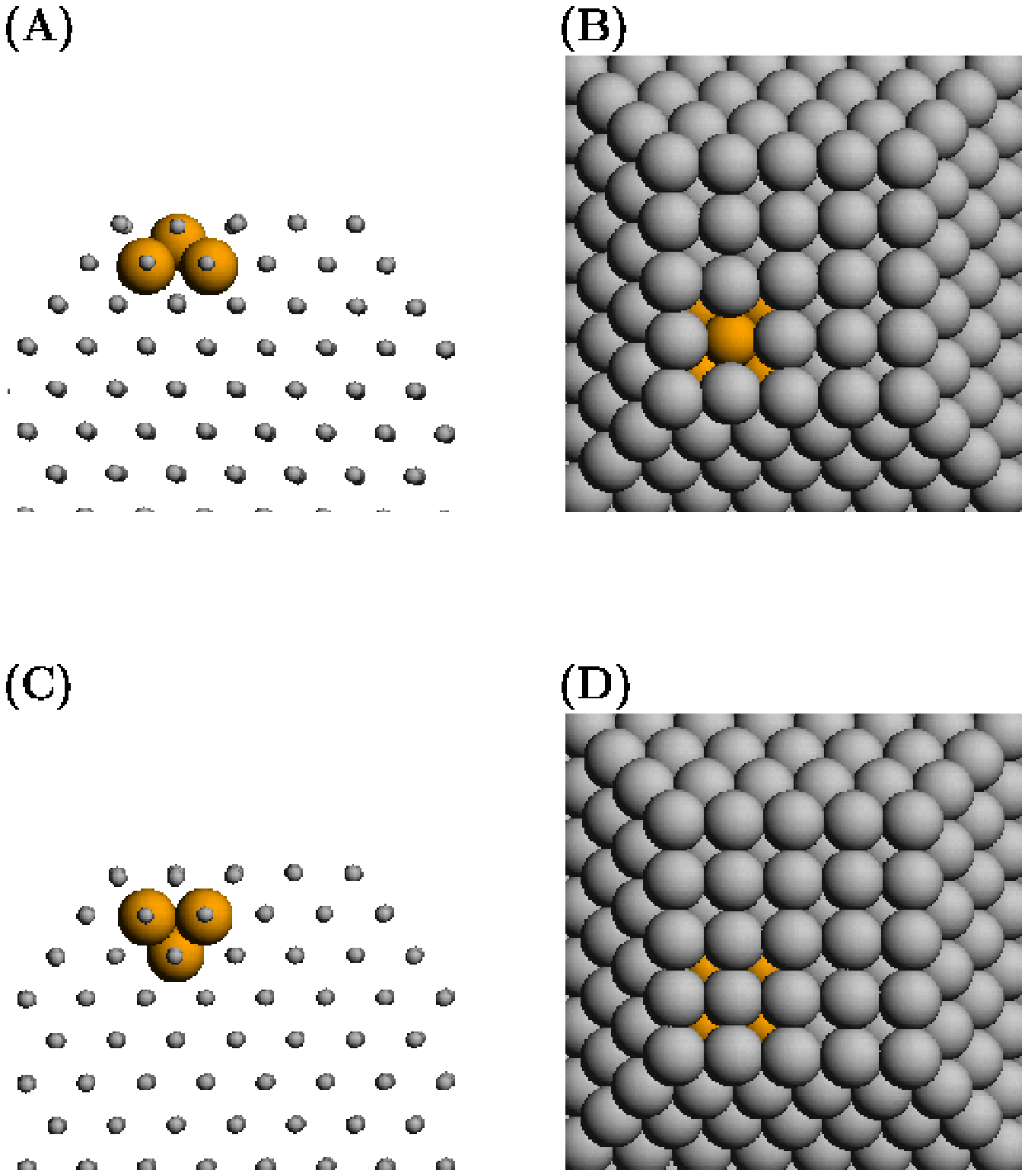}
 \caption{\label{fig_fcc_5atom} (A) and (B): side and top view of the lowest-energy structure of a cluster of 5 C atoms in a truncated octahedron of 1289 atoms, in the cases of AgCu, AgNi and AgCo. One C atom of the 5-atom pyramid appears at the cluster surface, in a (100) facet. (C) and (D): side and top view of the 5-atom pyramid obtained by putting its vertex inside the cluster, so that no C atom appears at the surface. In (A) and (C) S atoms are shown as smaller spheres}
\end{figure}

For inner and subsurface sites, in which the nearest-neighbor bond numbers are always the same, the results in Table \ref{tablefcc} show that the energetics is in general well correlated with the local pressure $P$ acting on the C atom. $P$ is always negative, indicating the expected tensile strain for impurities that are hosted in a matrix of bigger atoms. The most favorable site has the lowest absolute value of $P$ in all systems, indicating that the lowering of strain is an important driving force in selecting the most favorable sites. For what concerns other inner sites, the correlation between energy and low absolute pressure is perfect for AgCu, AgNi and AgCo, while there is some discrepancy in AuCo for what concerns the pressures of the second- and third-best sites.

For two C atoms, the most favourable configuration is always a dimer with one atom placed at a subsurface vertex position. Configurations with the two C atoms in dissociated configurations (for example, the C atoms placed at two different subsurface vertices) are always slightly higher in energy, by a minimum of 0.011 eV in AuCo to a maximum of 0.037 eV in AgNi.

The case of five C atoms is quite interesting because of the possible appearance of a C atom at the cluster surface. In fact, for AgCu, AgNi and AgCo the lowest-energy configuration contains a square-basis pyramid of C atoms which is placed as shown in Fig. \ref{fig_fcc_5atom}(A)-(B). The vertex of the pyramid appears at the cluster surface, on a (100) facet. The origin of this effect can be qualitatively explained as follows. Let us compare this pyramid with the inverted pyramid in which the vertex atom is exchanged with an Ag atom and placed inside the cluster, as in Fig. \ref{fig_fcc_5atom}(C)-(D). 

Therefore, the transformation from the inverted pyramid to the pyramid with the vertex at the surface amounts in total to bringing a C atom (the vertex of the pyramid) from an inner site to the (100) facet, which should have an energy cost in the range 0.10-0.15 eV for AgCu, AgNi and AgCo (see the discussion about a single C atom). 
However, if one calculates the local pressure $P$ on each atom of the pyramids, one obtains that its absolute values are well above 5 GPa for the inverted pyramid, while they drop down of about a factor 5 in the pyramid with the vertex at the surface. This better local relaxation overcompensates for the energy cost associated to bringing the C atom at the surface. The situation is different in AuCo, where this energy cost is much higher (about 0.3 eV) so that a better relaxation is not sufficient. In AuCo, the lowest-energy configuration has no C atoms at the cluster surface.

\begin{figure} [ht]
\includegraphics[width=11cm]{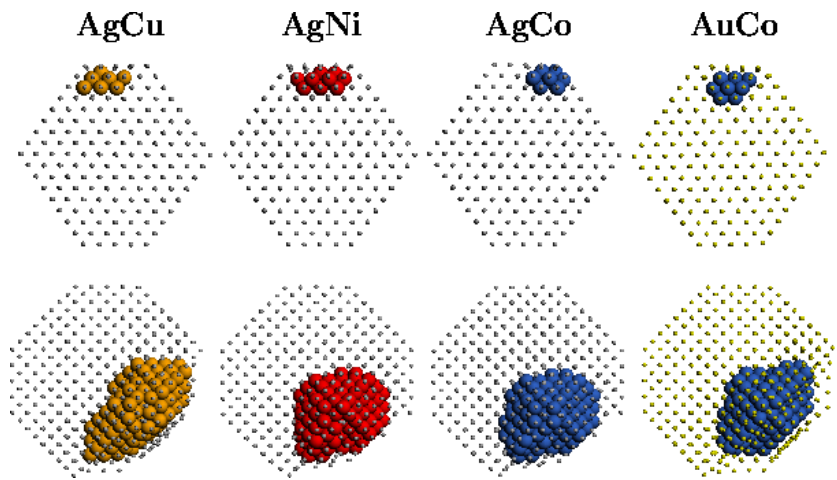}
 \caption{\label{fig_to1289_n} Optimized structures of fcc truncated octahedra of 1289 atoms, with composition S$_{1269}$C$_{20}$ (top row) and S$_{1089}$C$_{200}$ (bottom row). S atoms are shown as small spheres so that C atoms (bigger spheres) are visible.}
\end{figure}

For 20 C atoms the core is compact and off-center, with most of its atoms in the subsurface layer (see Fig. \ref{fig_to1289_n}). In AgCu and AgNi there are two atoms in the surface layer, whereas there are 8 in AgCo, and none in AuCo. The results found for AgCo is related to the low energetic cost related to placing Co atoms at  the cluster surface.

\begin{figure} [ht]
\includegraphics[width=11cm]{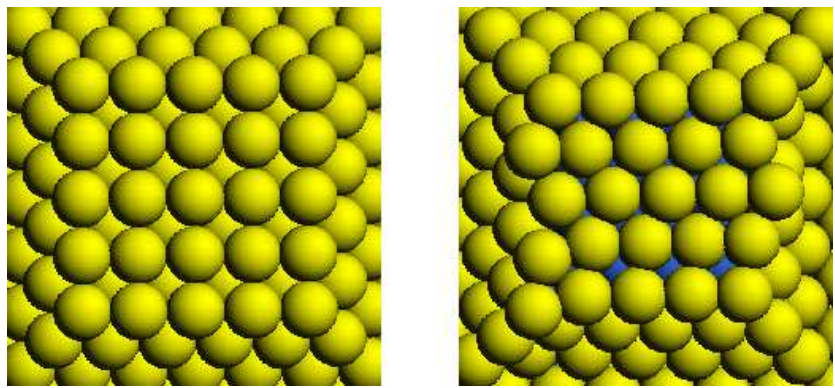}
 \caption{\label{fig_to1289_facet} Left: unreconstructed (100) facet of the 1289-atom AuCo truncated octahedron. Right: reconstructed (100) facet with Au atoms covering the underlying Co core.}
\end{figure}

In Fig. \ref{fig_to1289_n} we show also the optimized structures of truncated octahedra of 1289 atoms containing cores of 200 atoms. Again, the cores are clearly off-center. In all cases, part of the core is covered by a single Ag or Au monolayer.
Even though all systems present qualitatively similar behaviors, some differences can be noted. AgCo and AgNi present more compact cores, followed by AuCo and AgCu, which is the least compact. Moreover, in AgNi, and even more in AgCo, some C atoms appear at the cluster surface, especially in (001) facets, confirming the trends shown for smaller cores. These findings are in qualitative agreement with the fact that AgCo and AgNi have the strongest tendency against mixing \cite{Hultgren}, and a large difference in cohesive energy with Ag. Therefore the Co and Ni cores in Ag try to maximize the number of C-C nearest-neighbor bonds. We note however that the number of Co or Ni atoms at the surface remains comparatively small 
($\sim$10-15), because their appearance is mostly on the small (100) facets. This issue however deserves further investigation by fully unconstrained global optimization in which both cluster geometry and shape are optimized together \cite{Rossi09jpcm}, so that shell atoms are allowed to better find their optimal configuration. Work is in progress on this point.  

An interesting effect appears on the (100) facet of AuCo nanoparticle which is on the side where the core is placed. The Au atoms cover the Co core with a single layer which undergoes a reconstruction (see Fig. \ref{fig_to1289_facet}) of the same type which occurs on Au(100) bulk crystal surfaces \cite{Mochrie1990prl,Wang1991prl}. This reconstruction leads to the formation of a more compact surface layer, with a hexagonal pattern instead of a square pattern. At variance with the case of infinite surfaces, we find this reconstruction to occur only on the facets that cover the subsurface Co core, but not on other facets. The size mismatch between Au and Co atoms favors indeed Au atoms adopting more compact configurations, as those obtained by means of this reconstruction. This has been verified by artificially increasing the size of Co atoms $r_{0CC}$ while leaving all other parameters of the interaction potential unchanged. With no increase of $r_{0CC}$, the configuration with unreconstructed facet is a local 
minimum, even though it is somewhat distorted and 1.8 eV higher in energy than the reconstructed facet. If $r_{0CC}$ is increased by 2\%, the unreconstructed facet is still higher in energy (by 0.6 eV), but it is not distorted. A reconstructed facet in which a single Au row is displaced becomes slightly lower in energy than the configuration reported in the right panel of Fig. \ref{fig_to1289_facet}. For an increase of 5\%, the unreconstructed facet becomes lower in energy than the reconstructed one, and finally, for an increase of 10\% (at which C atoms are still smaller than S atoms), the reconstructed configuration is not even a locally stable minimum. The same tendency towards reconstruction is seen to a lesser extent in the AgCu case, confirming that an important driving force is indeed size mismatch, because Ag(100) bulk surfaces do not reconstruct.     

We note that in the reconstructed facet there are some Au atoms at the extremities of the rows that have rather low coordination, having five first neighbors only. This implies that displacing these atoms to other terrace positions has a small energy cost, of a few hundredths eV. It is thus expected that, with increasing temperature, these atoms may leave their original position to diffuse on the cluster surface. However, the facet should keep its reconstructed shape (with some shorter Au rows) because the unreconstructed geometry is much higher in energy.  

\begin{figure} [ht]
\includegraphics[width=11cm]{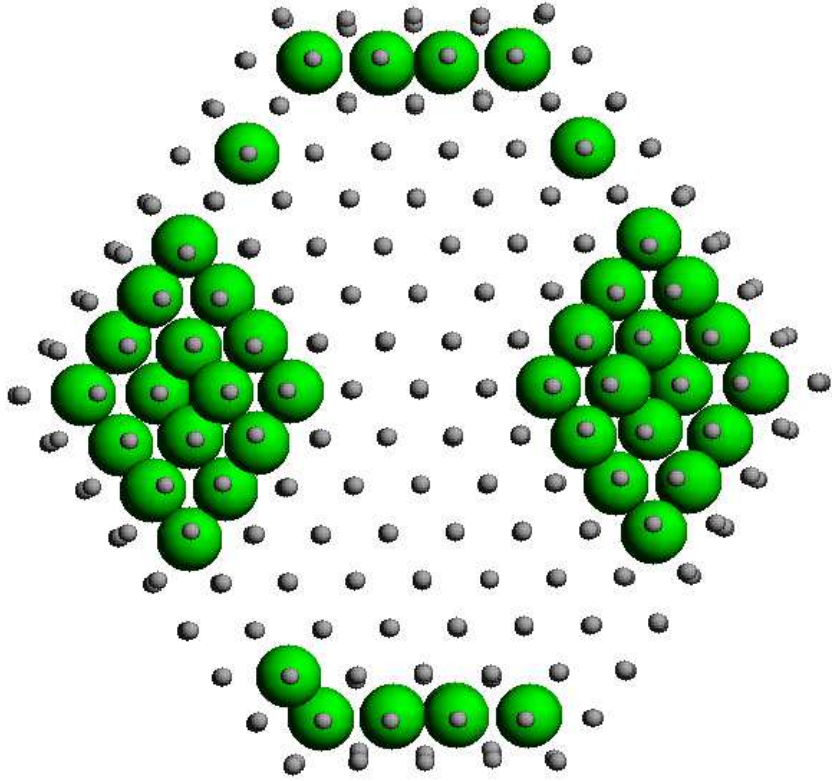}
 \caption{\label{fig_agcu_csi_5pc} Lowest energy structure of  Ag$_{1189}$Cu$_{100}$ for the case in which the mixed Ag-Cu interactions are artificially reinforced by increasing the parameter $\xi_{SC}$ by 5\%. Cu atoms fill subsurface positions below (100) facets instead of forming compact aggregates as those found the case in which the mixed interactions are not artificially reinforced.}
\end{figure}

In all systems we have found a clear preference in favor of compact cores. However, Monte Carlo simulations of an atomistic model of AuPt were revealing the possibility of finding three-shell onionlike structures as the equilibrium ones \cite{Wang2012jpcc}. In our case, these structures, having non-compact aggregates of C atoms, are metastable \cite{Baletto03prl}. In order to investigate the origin of the difference in behavior with respect to AuPt, we have verified whether it is possible to obtain onionlike structures in AgCu by artificially changing the interaction parameters. For example, we have increased  the parameter $\xi_{SC}$, which rules the strength of the attractive part of the mixed interactions, by 5\%, finding that non-compact aggregates of Cu atoms fill the subsurface shell, especially below vertices, edges and  (100) facets. These aggregates (see Fig. \ref{fig_agcu_csi_5pc}, reporting results for 100 C atoms) are very similar to those found for AuPt in Ref. \cite{Wang2012jpcc}. However, 
putting a larger number of Cu atoms leads 
to filling both subsurface sites below (111) facets and inner sites. Therefore, the perfect three-shell structure, with the surface shell being completely of Ag atoms, the subsurface shell of Cu atoms and the inner part of Ag atoms, is not obtained as the lowest-energy configuration, but as a metastable one \cite{Baletto03prl}. We note however that the surface shell is completely of Ag atoms.

\subsection{Icosahedra}

Icosahedral core-shell nanoparticles present an even richer behavior, with the onset of a new kind of morphological instability. Before showing how this instability occurs, a few words about icosahedral structures in general are necessary.
The Mackay icosahedron  \cite{Northby} has a shell structure. An icosahedron with $n$ shells ($n \geq 1$) has a number of atoms $N= (10 n^3 - 15 n^2 +11 n -3)/3$, which gives the series of magic numbers 1, 13, 55,  147, 309,  561,  923, 1415 .... In the following we consider compositions for which there is a number of C atoms that corresponds to the completion of $k$ shells, with $k < n$.

At variance with fcc and decahedra, the icosahedron is a structure whose central site is highly compressed. For monometallic clusters, the local pressure $P$ on the central site is positive and quite large (several ten GPa) so that eliminating the central atom and leaving a vacancy at its place can be energetically favorable \cite{Mottet97ss}, if the icosahedron is sufficiently large. For the same reason, introducing a smaller atom at the center of the icosahedron instead of the vacancy, as in the case of Cu, Ni, Co in Ag or Au, allows a notable strain release \cite{Mottet2005prl}. The local pressure remains still positive (compressive strain), even if the impurity atom is hosted in a matrix of larger atoms, but strongly decreases compared to the case in which the center is occupied by an atom of the same species as the matrix. For the small impurity, the central site thus becomes by far the most favorable place \cite{Baletto03prl}. This specific feature of the icosahedron will have important consequences 
for the structures of the cores, as we show below.

\begin{figure} [ht]
\includegraphics[width=11cm]{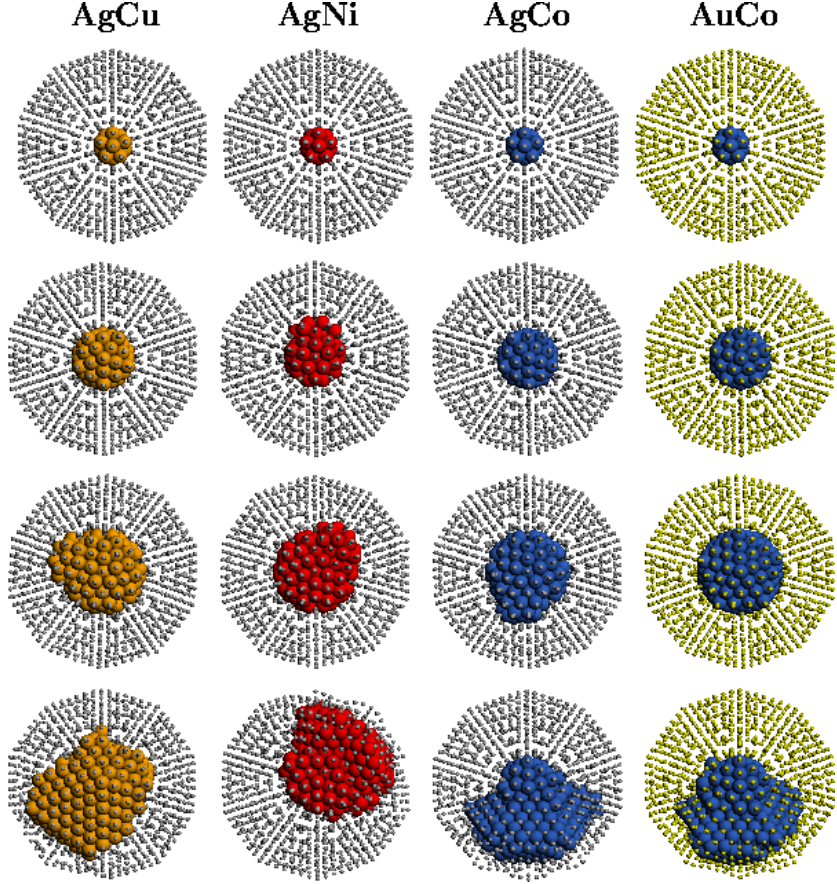}
 \caption{\label{fig_ih1415} Lowest-energy configurations of icosahedral nanoparticles of fixed size (number of shells $n=8$, total number of atoms $N=1415$) for increasing size of the core. From top to bottom: core of 13 atoms (number of shells $k=2$), of 55 atoms ($k=3$), of 147 atoms ($k=4$), of 309 atoms ($k=5$). S atoms are shown as small spheres so that C atoms (bigger spheres) are visible.}
\end{figure}

\begin{figure} [ht]
\includegraphics[width=11cm]{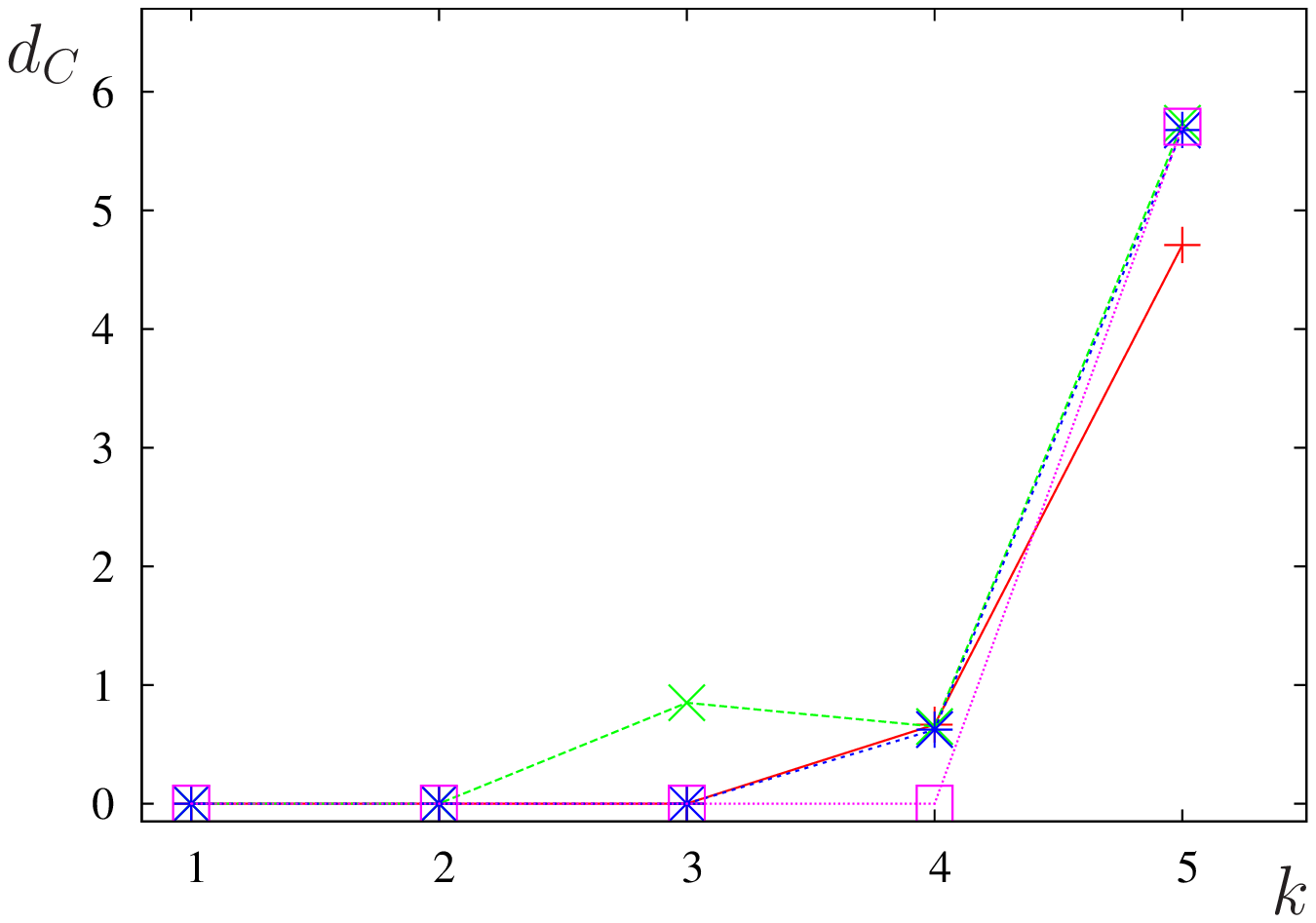}
 \caption{\label{fig_spostamento} Displacement $d_C$ (in \AA) of the geometric center of the core with respect to the center of the icosahedron, as a function of the number of shells in the core $k$. Different symbols refer to different systems ($+$ to AgCu, $\times$ to AgNi, $*$ to AgCo, and $\square$ to AuCo). }
\end{figure}

As a first case we consider an icosahedron with fixed $n=8$, i.e. with 1415 atoms in total, and increase the number of C atoms  (we consider 1, 13, 55, 147, 309 C atoms, corresponding to $k$ from 1 to 5, as shown in Fig. \ref{fig_ih1415}). For $k=1$ (not shown in the figure), the single C atom is always preferentially placed at the center of the structure as expected. For $k=2$, the 13 C atoms are again placed at the nanoparticle center in all cases. However, when $k$ increases above 2, a different behavior appears in AgNi. The 55 Ni atoms are not all contained in the inner shells with $k \leq 3$, but some of them are now placed in the 4$^{th}$ shell, with some Ag atoms being correspondingly in the 3$^{rd}$ shell.  For AgCu, AgCo, and AuCo, the C 
atoms still form a perfectly centered icosahedral core.

When $k$ increases to 4 and 5, the morphological instability of the cores becomes more and more evident, with the development of irregular shapes that extend asymmetrically from the center towards the surface of the nanoparticle, filling some parts of the 6-th and 7-th shells. 
The asymmetric core however does not extend to the 8-th shell because C atoms gain in occupying subsurface positions, as we have seen for fcc clusters \cite{Baletto03prl,Langlois2012nanoscale}, so that none or very few C atoms appear at the cluster surface. The displacement of the core from the symmetric position is quantified in Fig. \ref{fig_spostamento} by measuring the distance $d_C$ of the core geometric center from the central site of the icosahedron. The instability begins when  $d_C$ shows a non-zero value. One can distinguish a \textit{weak} form of morphological instability, when only sites of the $(k+1)^{th}$ shell are occupied by C atoms and $d_C$ has a small value, from a fully developed form, which is associated to  much larger values of $d_C$. 

\begin{table} [ht]
 \begin{tabular}{l l |cccc}  
    &     &    AgCu    &    AgNi    &    AgCo    &    AuCo   \\
  n &  k  & $\Delta E$ & $\Delta E$ & $\Delta E$ & $\Delta E$\\
  \hline
  8 &  1  &    0.00    &   0.00     &   0.00     &     0.00  \\
  8 &  2  &    0.00    &   0.00     &   0.00     &     0.00  \\
  8 &  3  &    0.00    &  -0.07     &   0.00     &     0.00  \\
  8 &  4  &   -0.52    &  -2.06     &  -0.89     &     0.00  \\
  8 &  5  &   -4.34    & -14.85     & -13.55     &    -7.05  \\
  6 &  4  &   -1.53    &    -       &     -      &    -2.69  \\
  7 &  4  &   -0.85    &    -       &     -      &    -0.40  \\
\hline  
\end{tabular}
\caption{\label{tabledeltae} Energy differences $\Delta E$ (in eV) between the lowest energy structure and the structure with centered symmetric core. Negative $\Delta E$ indicate that asymmetric cores are more favorable.}
\end{table}

We remark that centered symmetric core shapes still remain locally stable minima at least up to $k=5$ in all cases. However, with increasing $k$, the energy difference between the most favorable asymmetric core and the metastable centered symmetric core increases, becoming very large for $k=5$ especially for AgNi and AgCo (see Table \ref{tabledeltae}).

These results show that there is a critical value of $k$, referred to as $k_c$, at which the morphological instability starts. $k_c$ is system-dependent. For $n=8$,  $k_c=3$ for AgNi, $k_c=4$ for AgCu and AgCo, $k_c=5$ for AuCo.

\begin{figure} [ht]
\includegraphics[width=11cm]{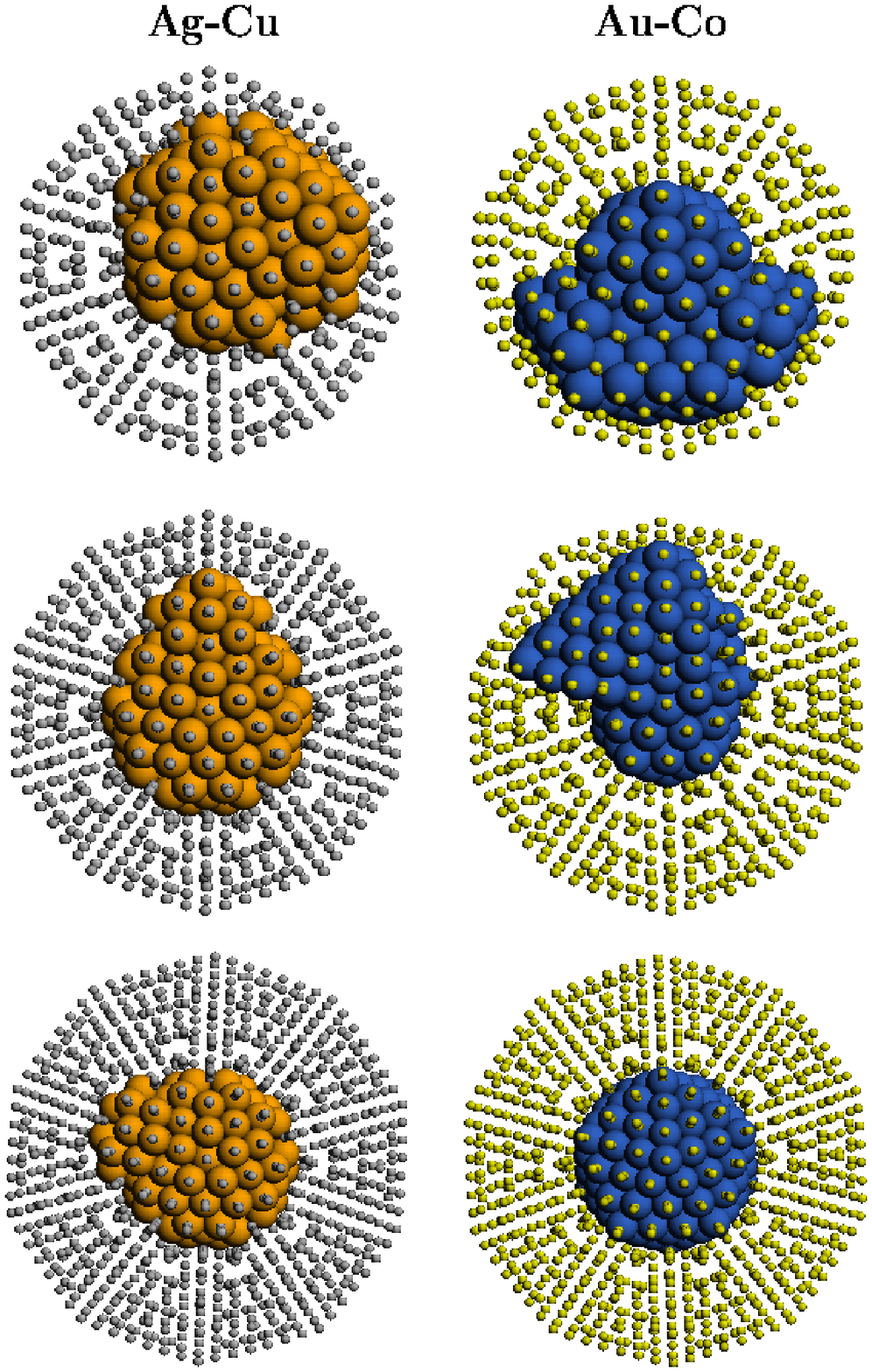}
 \caption{\label{fig_ih_core147} Lowest-energy configurations of icosahedral nanoparticles of fixed core size (number of shells $k=4$, corresponding to 147 C atoms) for increasing number of S atoms (and size of the nanoparticle). From top to bottom: $N=561$ (number of shells $n=6$), $N=923$ ($n=7$), $N=1415$ ($n=8$). S atoms are shown as small spheres so that C atoms (bigger spheres) are visible.}
\end{figure}

Now we consider the possible dependence of $k_c$ on $n$. In Fig. \ref{fig_ih_core147} we fix $k=4$ and increase $n$ from 6 to 8. The cores that are asymmetric for $n=6$ become more and more centered and symmetric for $n=7$ and 8, showing that increasing the shell thickness tends to stabilize the morphology of the cores. This shows for example that, in AuCo, $k_c=4$ for $n=6$ but it increases to $k_c=5$ for $n=8$.

What is the physical origin of this morphological instability? The instability has the same qualitative features in all systems, with some quantitative differences, developing earlier in AgNi, then in AgCu and AgCo and finally in AuCo, the latter system being able to support the largest centered symmetric cores. The onset of the instability in all these systems suggests that there might be an explanation of generic character, originating from some simple common feature, and being possibly applicable to an even wider class of systems.

The key property that has to be considered turns out to be the lattice mismatch between the elements in the nanoparticle. This can be understood by quite simple considerations. In the bulk crystals, the nearest neighbor distances between S and C atoms are $r_{0S}$ and $r_{0C}$, respectively. Now we change artificially $r_{0C}$ in our model letting all other parameters of the model unchanged. This amounts to changing the size of C atoms only. Then we look for the value $r_{0C}^*$ which minimizes the energy of the nanoparticle, and we compare it to the original $r_{0C}$ \cite{Rapallo05jcp}.  We repeat this procedure for icosahedra of 1415 atoms ($n=8$) with centered symmetric icosahedral cores with different $k$, from 1 to 5. 

Let us consider for example the case of AgNi, for which $r_{0S}= 2.89\,$\AA \; and $r_{0C}= 2.49\,$\AA. When $k=1$ the single Ni atom in the central site has an optimal size $r_{0C}^*$ which is smaller than $r_{0C}$ by 8.3\%. This means that the strain release which is achieved by placing a Ni atom at the center of an Ag icosahedron would be even more effective if the size of the Ni atom would be smaller than its actual size. The situation begins to change for $k=2$. The 13 Ni atoms form a small central icosahedron. However, in this case $r_{0C}^*$ is slightly larger than $r_{0C}$, by 2.7\%, so that the best strain release would be achieved if Ni atoms were slightly bigger. This tendency becomes more evident for $k=3$, for which $r_{0C}^*$ is larger than $r_{0C}$ by 5.7\%. This means that the actual size of Ni atoms  is indeed too small to support a centered icosahedral core of 55 atoms . This core is therefore too small, so that it may benefit by the substitution of some of its Ni atoms by larger Ag atoms, 
as shown in Fig. \ref{fig_ih1415}, second row. The behavior of AgCu and AgCo is of the same type, but the instability occurs at larger core sizes due to a smaller lattice mismatch between the constituent elements. The same line of reasoning apply to AuCo too, even though the comparison with the other systems is less direct due to the fact S atoms are Au.

\begin{figure} [ht]
\includegraphics[width=11cm]{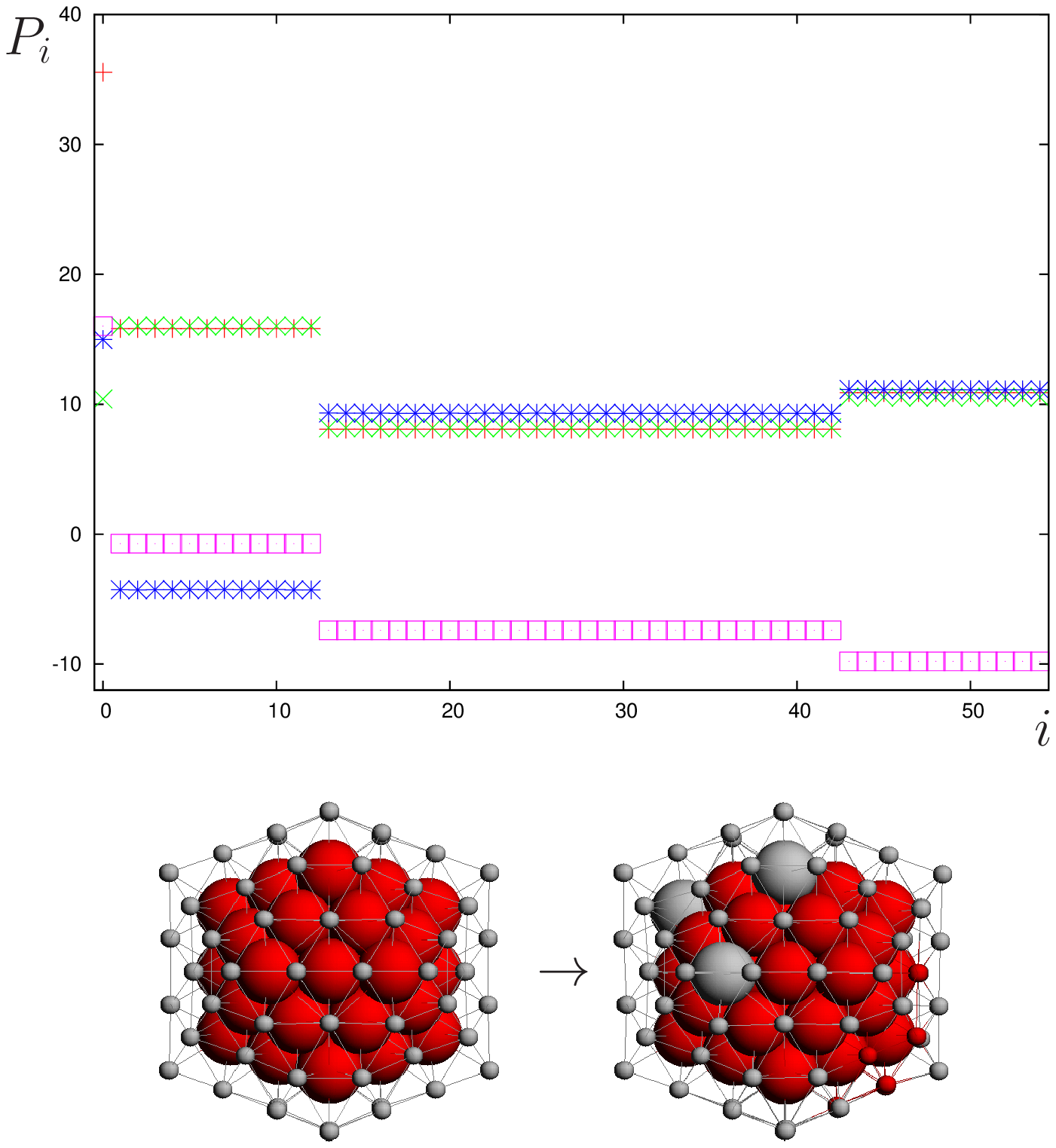}
 \caption{\label{fig_agni1415_pressure} Local pressure $P_i$ in an icosahedron of 1415 atoms, for its inner 55 atoms that are arranged with increasing distance from the center. Different symbols refer to different compositions, i.e. different number of Ni atoms. $+$: pure Ag cluster ($k=0$). $\times$: single Ni impurity at the central site ($k=1$). $*$: centered icosahedron of 13 Ni atoms ($k=2$). $\square$: centered icosahedron of 55 Ni atoms ($k=3$). The bottom pictures show the difference between the centered symmetric (left) and the optimal (right) cores in the case of 55 Ni atoms. In the optimal core, some Ag atoms substitute the Ni atoms at vertex positions in the third shell. These Ni atoms are now placed in the fourth shell, which is shown by smaller spheres.}
\end{figure}

The physical effect triggering the instability can be analyzed also by calculating the local pressure $P_i$ acting on each atom, as given by Eqs. (\ref{pressure}-\ref{pressurefin}).

In Fig. \ref{fig_agni1415_pressure} we report $P_i$ for the inner 55 atoms for the 1415-atom icosahedron in the case of AgNi with centered symmetric icosahedral cores. For pure Ag ($k=0$) $P_i$ is strongly positive at the central site, which is therefore strongly compressed, and, on average decreases with the distance from the center \cite{Marks1984}, being however positive for all atoms belonging to the first three shells. Substituting a single Ni impurity decreases $P_i$ at the central site \cite{Mottet2005prl}, but the local pressure in other sites remains nearly unchanged.
For $k=2$ we have two shells of Ni atoms, and we begin to see some qualitative change in the behavior of $P_i$. In fact, the pressure of the 12 Ni atoms in the second shell becomes  negative, indicating again that these atoms show some tendency to expand. Finally, for $k=3$,  $P_i$ becomes more strongly negative for the atoms of the third shell, especially for those that are at vertex sites of the 55-atom icosahedron. In this case it is favorable  to substitute these vertex atoms with bigger atoms, thus obtaining a small energy gain. In the lowest-energy structures we have thus a few Ag atoms in the third shell in vertex position and, correspondingly, the same number of Ni atoms is now forming a small island in the fourth shell.

This morphological instability is thus originating from the accumulation of strain as the number of shells in the core increases. The key point is that strain changes from a compression for small cores to a stretching for large cores. Since S atoms are of bigger size, this kind of strain can be somewhat released by substitution of C atoms in the inner shells. This leads to an energy gain, which is small for the first substitutions but becomes quite large as the number of C atoms increases, as shown in Table \ref{tabledeltae}. 

The morphological instability of the cores therefore bears some common points with the Stranski-Krastanov (SK) instability occurring in heteroepitaxial thin film growth \cite{Markov}. In heteroepitaxial growth, atoms of species B are deposited on a substrate of species A. Between the two species a lattice mismatch is present. As the number of deposited layers increases, strain accumulates so that the film breaks up in its top part, the growth mode changing from layer-by-layer to three-dimensional, with the formation of mounds. The SK instability occurs when there is a (free) energy gain in starting the $(n+1)^{th}$ layer before completing the $n^{th}$ layer \cite{Markov}.
Therefore both SK instability and core instability in icosahedra share a common origin. There are however some differences. 

In the SK instability, the strain corresponds to either compression or stretching, depending on the sign of the lattice mismatch. In the core instability, the strain changes sign,  from the compression, which is specifically inherent to the icosahedral geometry, to the stretching which originates by the increases of the number of shells in the core. Therefore the core instability requires a specific sign of the lattice mismatch, with C atoms being smaller than S atoms.

Moreover, in the SK instability, the strain is caused by the lattice mismatch at the bottom layer of the thin film, and it is released by breaking the thin film in its outer layers. In the core instability, strain is originated by lattice mismatch at the outer shell of the core, and it is released by breaking these outer shells.  
We note that concerning the critical core size, we have determined a critical number of shells $k_c$ for its onset. More precisely, the instability starts from a number of C atoms $N_C$ which is between the numbers of atoms necessary to complete the shell with $k_c -1$ and the shell with $k_c$.

We remark that the term \textit{morphological instability} does not mean that centered cores are unstable configurations. They are locally stable minima, whose energy is however much higher than the energy of the optimal configuration. The term morphological instability means that the centered morphology does not persist as the lowest-energy morphology as the proportion of C atoms increases in the cluster, i.e. the term instability is referred to the overall core morphology and not to the local minimum of the centered core. 

We briefly discuss now what happens going towards C-rich compositions. For example, we have considered the case $k=n-1$ in AgCu. For Cu-rich cases, it was previously shown that the Mackay structure of the shell is not at all the optimal one, anti-Mackay or chiral shells being much lower in energy \cite{Bochicchio2010nl,Bochicchio2012epjd}. Essentially, the Mackay shell contains too many atoms to well adapt itself to a ''small'' substrate. Anti-Mackay or chiral shells contain fewer atoms so that they can adapt much better. If the number of Ag atoms is that of a Mackay shell, the optimal configuration is indeed the anti-Mackay or chiral shell with the Ag atoms in excess forming an island above it. Therefore it is now the shell which loses its symmetry, while the core can stay symmetric. However it is possible to build a Mackay shell for $k=n-1$, and then optimize chemical ordering by exchanges only, thus keeping this geometric structure. This leads to an intermixing of Ag and Cu within the surface shell (
which is heavily strained if made of Ag only), with some Ag atoms correspondingly going inside the cluster.

Finally consider icosahedra with a central atomic vacancy \cite{Mottet97ss}, that might be more stable even in the bimetallic case especially at large sizes. We consider in details AgCu with $n=8$, corresponding to a cluster of 1414 atoms. It turns out that the behavior is quite the same as in the case without the vacancy. For $k=2$ and 3 (12 and 54 Cu atoms respectively)  we find  centered symmetric icosahedral core morphologies with a central atomic vacancy. From $k=4$ on, the morphological instability takes place, as in the case of no vacancy.

\subsection{Decahedra}

The most favorable type of decahedron for pure Ag and Au clusters of sizes in the range $\sim 10^3$ atoms is expected to be the Marks decahedron \cite{Marks94}. Decahedral clusters have a single fivefold axis, which is shared by the five tetrahedra composing the structure. The axis terminates into two fivefold vertices. Most part of the decahedron surface is made of (111)-like facets, but there are five (100)-like open facets too. In the following we focus on a Marks truncated decahedron of 1228 atoms.

\begin{figure} [ht]
\includegraphics[width=11cm]{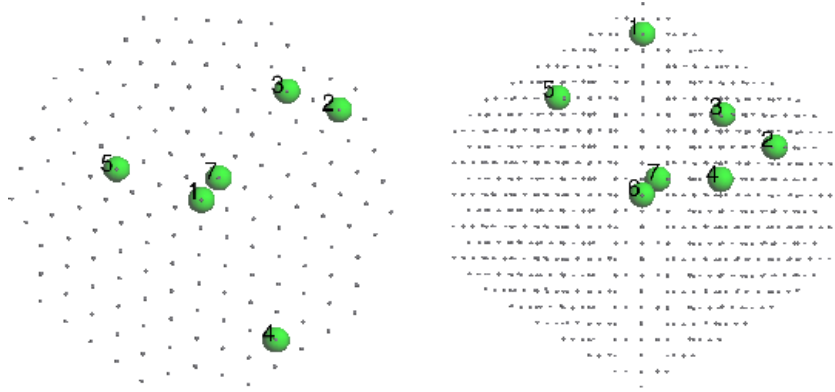}
 \caption{\label{fig_dh1228_imp} Single-impurity sites in a decahedron of 1228 atoms considered in Table \ref{tabledh}.}
\end{figure}

\begin{table}   [ht]
 \begin{tabular}{l|c|c|c|c|c|c|c|c}
         &  AgCu & AgCu & AgNi &  AgNi &  AgCo &  AgCo & AuCo & AuCo  \\
position &  $E$  & $P$  & $E$  &  $P$  &   $E$ &   $P$ &  $E$ &  $P$  \\
  \hline
1 subsurface, below a fivefold vertex            & -0.063  &  -4.47 & -0.137  &  -8.23  & -0.169 &  -4.61  & -0.161  &  -7.46 \\
2 subsurface, below the edge of a (100) facet    &  0.013  &  -6.49 & -0.025  & -10.16  & -0.068 &  -7.13  & -0.041  &  -9.54 \\
3 subsurface, below the edge of a (111) facet    &  0.032  &  -7.81 &  0.012  & -11.44  & -0.012 &  -8.46  &  0.002  & -11.77 \\
4 subsurface, below the center of a (100) facet  &  0.054  &  -8,30 &  0.052  & -12.00  &  0.022 &  -9.17  &  0.099  & -12.78 \\
5 subsurface, below the center of a (111) facet  &  0.060  &  -9.22 &  0.067  & -13.10  &  0.051 & -10.49  &  0.096  & -13.89 \\
6 central, on the fivefold axis                  &  0.000  &  -7.94 &  0.000  & -11.60  &  0.000 &  -8.78  &  0.000  & -13.42 \\
7 central, nearest neighbor of the fivefold axis &  0.038  &  -9.36 &  0.055  & -13.13  &  0.053 & -10.66  &  0.063  & -14.87 \\
\hline  
\end{tabular}
\caption{\label{tabledh} Energy $E$ and local pressure $P$ of a single impurity in some different sites in a decahedron of 1228 atoms, after local relaxation. The sites are represented in Fig. \ref{fig_dh1228_imp}. The energy in the central position inside the cluster is set to zero in all systems. Energies are given in eV, local pressures in GPa.}
\end{table}

By analyzing the local pressure in a pure metal Ag or Au decahedron, it turns out that the sites with the highest local positive pressure are subsurface sites below the fivefold vertices, which thus belong to the fivefold axis. Their local pressure is not so high as in the central site of the pure-metal icosahedron, but it is still notable (8 and 16 GPa in the Ag and Au decahedron, respectively). It is therefore not surprising that these are the most favorable sites for small impurities, as shown in Table \ref{tabledh} and Fig. \ref{fig_dh1228_imp}. Other favorable sites are either along the fivefold axis or in subsurface positions, especially below (100)-like facets. 

\begin{figure} [ht]
\includegraphics[width=11cm]{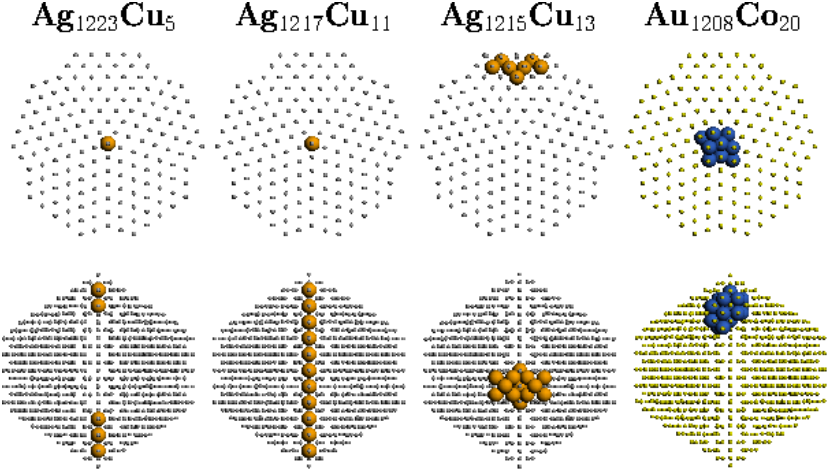}
 \caption{\label{fig_dh1228_small} Optimized structures of Marks decahedra of 1228 atoms and different compositions. Each cluster is shown in top and side view (first and second row, respectively). S atoms are shown as small spheres so that C atoms (bigger spheres) are visible.}
\end{figure}

\begin{figure} [ht]
\includegraphics[width=11cm]{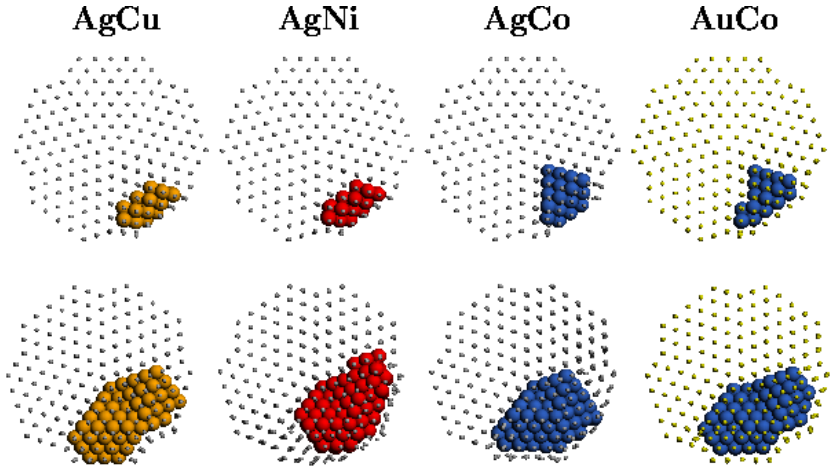}
 \caption{\label{fig_dh1228_n} Optimized structures of Marks decahedra of 1228 atoms, with compositions S$_{1178}$C$_{50}$ (top row) and S$_{1028}$C$_{200}$ (bottom row). S atoms are shown as small spheres so that C atoms (bigger spheres) are visible.}
\end{figure}

Favorable sites along the fivefold axis are arranged along a straight line, therefore it is not possible to build up compact aggregates by filling them only. On the other hand, this is possible by filling subsurface sites below (100) facets. This generates an interesting competition at small core sizes between the tendency to fill the most favorable sites and the tendency to form compact aggregates. This competition is especially evident in AgCu. For Ag$_{1226}$Cu$_{2}$, the two subsurface sites below vertices are occupied. Increasing the number of Cu atoms to obtain  Ag$_{1223}$Cu$_{5}$, two small linear aggregates (a dimer and a trimer) are formed along the fivefold axis as shown in Fig. \ref{fig_dh1228_small}.
The fivefold axis is filled by Cu atoms up to composition Ag$_{1217}$Cu$_{11}$, at which all inner sites of the axis are filled. Up to this composition, the tendency to fill the most favorable sites prevails over the tendency to form compact aggregates.
However, a further increase of Cu causes the tendency in favor of compact aggregates to prevail. At composition Ag$_{1215}$Cu$_{13}$, a transition to completely different core shapes, that are more compact and below (100) facets, finally takes place.
In AgNi and AgCo, the behavior is of the same kind, but the transition to compact shapes occurs at even smaller core sizes. 

On the other hand, AuCo shows a different behavior. In fact, the small compact cores do not form below a (100)-like facet, but in subsurface position below a fivefold vertex, as shown in Fig. \ref{fig_dh1228_small} for Au$_{1208}$Co$_{20}$. 

However, in all systems, large cores form below (100)-like facets as shown in Fig. \ref{fig_dh1228_n} for cores of 50 and 200 atoms. These cores are clearly off-center and somewhat resemble those found in fcc nanoparticles. The cores are slightly more compact for AgNi and AgCo. For AgCu and AuCo the most favorable core shapes of size 200 are quasi-symmetric with respect to a twin plane, resembling two incomplete tetrahedral units that share a few atoms on the twin plane.

\subsection{Polyicosahedra}

\begin{figure} [ht]
\includegraphics[width=11cm]{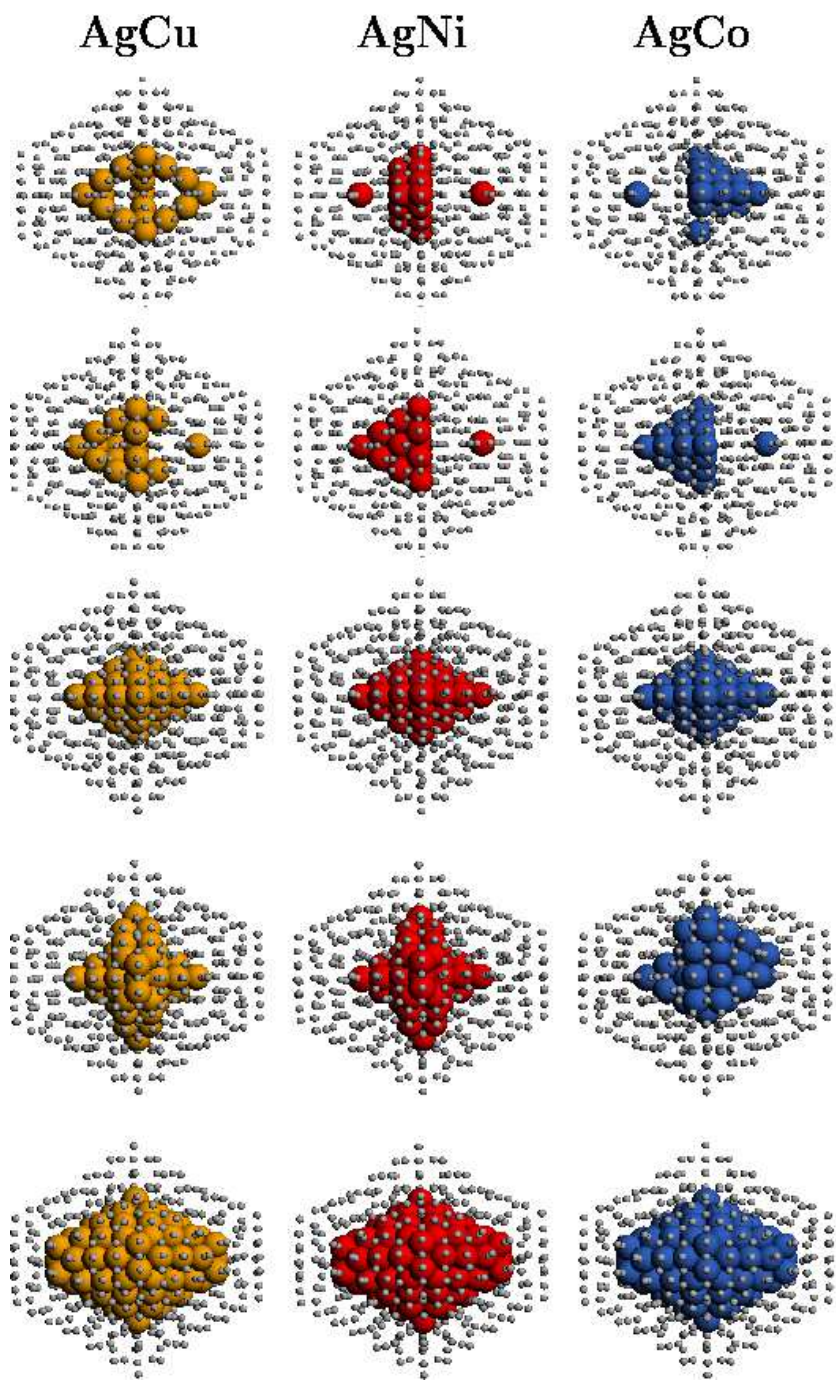}
 \caption{\label{fig_pentaih1} Optimized structures of polyicosahedra of 419 atoms. From top to bottom, the compositions are: S$_{401}$C$_{18}$, S$_{398}$C$_{21}$, S$_{389}$C$_{30}$, S$_{365}$C$_{54}$, and S$_{323}$C$_{96}$. S atoms are shown as small spheres so that C atoms (bigger spheres) are visible.}
\end{figure}

Polyicosahedra are made of several icosahedra that share some atoms. Core-shell polyicosahedra have been theoretically predicted \cite{Rossi04prl} as being very stable structures for AgNi and AgCu clusters for small sizes, below 50 atoms. For these sizes, the polyicosahedra are made of interpenetrating small icosahedra of 13 atoms, which in turn consist of a central atom (which is always either Cu or Ni) and of its first coordination shell, containing 12 atoms that can be in general of both species. However this is not the only way of building polyicosahedra. In fact, for larger sizes, one can consider larger icosahedra as elementary building blocks \cite{Bochicchio2010nl}. In the following, we will consider the Mackay icosahedron \cite{Northby} of 147 atoms as the elementary building block, that allows to build up clusters of a few hundred atoms. In particular, we deal with a pentaicosahedron of 419 atoms (see Fig. \ref{fig_geom}). If C atoms are few, this structure heavily distorts in the case of AuCo. 
Therefore we consider only AgCu, AgNi and AgCo in the following, unless otherwise specified. The results concerning the most stable structures are shown in Fig. \ref{fig_pentaih1}.

\begin{figure} [ht]
\includegraphics[width=11cm]{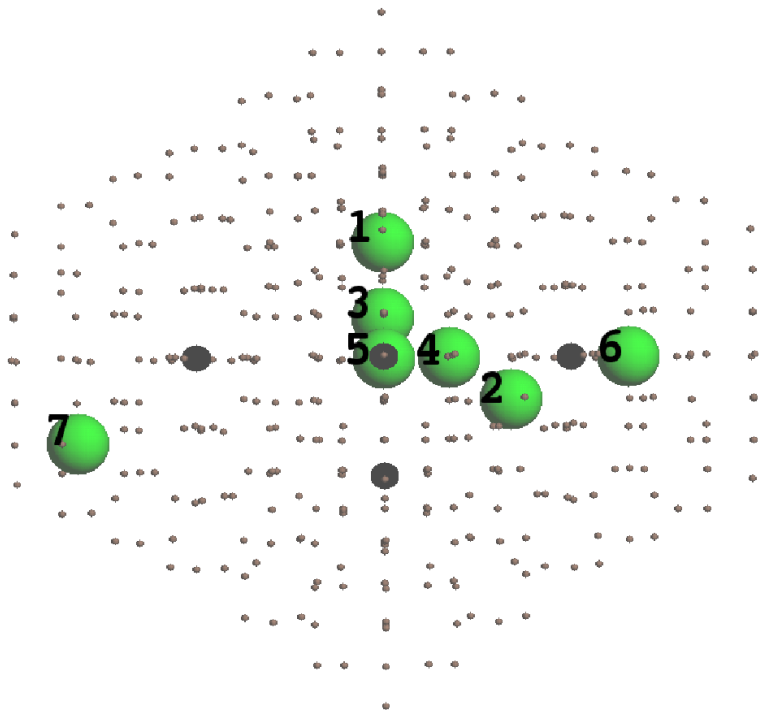}
 \caption{\label{fig_penta_num} Single-impurity sites in a polyicosahedron of 419 atoms considered in Table \ref{tablepenta}. Site 1 is an icosahedral center, sites 2 and 3 are on segments joining icosahedral centers, sites 4 and 6 are neighbors of an icosahedral center, site 5 is in the middle of the triangle whose vertices are icosahedral centers, site 7 is subsurface. The four icosahedral centers besides site 1 are indicated by mid-size spheres. Other sites are indicated by small spheres.}
\end{figure}

\begin{table}   [ht]
 \begin{tabular}{l|c|c|c|c|c|c}
         &  AgCu & AgCu & AgNi &  AgNi &  AgCo &  AgCo \\
position &  $E$  & $P$  & $E$  &  $P$  &   $E$ &   $P$ \\
  \hline
1 icosahedral center                            &  0.000  &  10.02 &  0.000  &   5.84  &  0.000 &  10.48  \\
2 segment joining icosahedral centers           &  0.381  &  -0.77 &  0.494  &  -4.86  &  0.467 &  -1.28  \\
3 segment joining icosahedral centers           &  0.410  &  -1.65 &  0.527  &  -5.69  &  0.505 &  -2.05  \\
4 neighbor of icosahedral center                &  0.477  &  -3.36 &  0.621  &  -8.07  &  0.574 &  -4.66  \\
5 center of the triangle of icosahedral centers &  0.526  &  -4.74 &  0.681  &  -8.60  &  0.648 &  -5.36  \\
6 neighbor of icosahedral center                &  0.569  &  -6.20 &  0.735  & -10.11  &  0.704 &  -6.97  \\
7 subsurface position                           &  0.583  &  -3.41 &  0.687  &  -7.76  &  0.562 &  -4.20  \\
\hline  
\end{tabular}
\caption{\label{tablepenta} Energy $E$ and local pressure $P$ of a single impurity in some different positions in a polyicosahedron of 419 atoms, atoms, after local relaxation. Energies are given in eV, local pressures in GPa. Site numbers refer to Fig. \ref{fig_penta_num}.}
\end{table}

The pentaicosahedron has five icosahedral central sites, that are separated from each other by two sites in between. Instead of forming a single aggregate, C atoms begin to populate the different centers, forming very small aggregates,  until there are atoms enough to merge in a single connected aggregate. As can be seen in the first row of Fig. \ref{fig_pentaih1}, the way by which the different small aggregates merge is system-dependent. In AgCu, a non-compact network-like aggregate forms, with Cu chains joining the icosahedral centers. In AgNi and AgCo, a more compact aggregate of two layers forms, comprising three icosahedral centers, plus two isolated atoms in the remaining icosahedral centers. For 21 C atoms (second row) AgCu continues the network-like grown sequence, whereas AgNi and AgCo form a 20-atom tetrahedron (comprising four icosahedral centers) plus an isolated atom in the fifth center. Finally, for 30 C atoms, all systems form the same structure, a single connected aggregate with shape of a 
double tetrahedron whose five vertices are the icosahedral centers. Increasing the number of C atoms further, the growth sequence for AgCu and AgNi is more symmetric, with a chain of atoms developing around the common basis of the two tetrahedra forming the double tetrahedron. For AgCo, the structures are less symmetric. However, in all cases, small 13-atom icosahedra of the C species form around the icosahedral centers. Finally, for 96 C atoms the growth sequence is completed by a cluster which preserves the symmetry of the original pentaicosahedron. Around each icosahedral center there is a small 13-atom icosahedron of C atoms. This last structure is stable also for AuCo.

The growth sequences of the cores can be understood by considering site energies of single impurities and their local pressures, as reported in Table \ref{tablepenta}, which refers to the sites shown in Fig. \ref{fig_penta_num}. The five icosahedral are by far the preferred sites. C atoms in these sites present a positive pressure, which is however by far smaller than the pressure that would act on an Ag atom in the same sites. Therefore, as in simple icosahedra, placing the impurity in a icosahedral center causes a notable strain release and the stabilization of the structure. The second-best sites for a single impurity are then those on the segments joining the icosahedral centers. This is the driving force towards the formation of network-like aggregates. This driving force is counterbalanced by the tendency to form compact aggregates, which originates from the fact that the C-C bonds are the strongest ones.
In AgCo and AgNi, where C-C bonds are by far stronger, the tendency towards more compact structures  prevails. In AgCu, C-C bonds are less strong, therefore the driving force towards the network-like structures prevails up to about 20 Cu atoms. We note that low absolute values of the local pressures are well correlated with favorable site energies.

\section{Conclusions}

Our results clearly show that the optimal shape and placement of the core in core-shell nanoparticles are strictly correlated to the overall geometry of the nanoparticle itself. However some common features of general character can be singled out. The first point is that off-center asymmetric cores are the rule as lowest-energy structures, not the exception. In fact, we have seen that these cores are dominant in crystalline and decahedral motifs. Centered cores can be obtained in icosahedral nanoparticles, but only for core sizes that are not too large. In fact, our calculations show that there is a critical size for the onset of a morphological instability which drives the core to an asymmetric shape extending towards the nanoparticle surface. In polyicosahedral nanoparticles (and to a much lesser extent in decahedra), multi-center core structures are possible. Even in these cases the cores are often not symmetrically placed with respect to the center of the whole structure.

In all cases in which off-center cores are prevailing, centered cores are still possible as metastable configurations, whose lifetime can drastically change depending on temperature. 

The physical factors causing the asymmetric placement of the cores and the onset of the morphological instability of the cores in icosahedra originate from very simple features of generic character, essentially from the strain originated from lattice mismatch. The morphological instability of icosahedral cores bears some analogy with a very general phenomenon taking place in thin film heteroepitaxial growth, i.e. the Stranski-Krastanov instability.  
For this reason, we believe that the same effects should appear in a whole class of nanoparticles sharing the same features, i.e. poor miscibility, lattice mismatch and tendency of bigger atoms to segregate to the surface. For example, systems as AuNi, AuRh and AuPt are expected to present this kind of behavior.

Even though our calculations are finding low-energy structures and therefore are related to the low-temperature behavior of the systems, we believe that the asymmetric cores should be present also in a significant temperature range. In fact, energy differences between centered cores and off-center cores become readily very large, so that we expect that the effect of temperature may change the details of the core shape, however without changing its qualitative features. On the other hand, the melting behavior of icosahedra with small centered cores compared to the melting of icosahedra with larger off-center cores is an interesting subject which deserves investigation. Also the magnetic properties in Co- or Ni-containing nanoparticles with off-center cores can be of interest. 

Finally we note that our results are relevant to the design of experimental procedures to achieve
effective coatings of the cores in systems in which lattice mismatch is present. Our results show indeed that achieving effective coatings may be a quite difficult task. In fact, in the lowest energy configuration, the core very often tends to approach the nanoparticle surface, so that part of the core itself is covered by a very thin layer of shell atoms. A few core atoms may even appear at the cluster surface in some cases. Centered cores can be experimentally grown, but these are metastable structures, which therefore may present aging problems due to this driving force towards more asymmetric shapes.  

\section{Acknowledgements}
The authors thank I. Atanasov, R. L. Johnston and  L. D. Marks for useful discussions. The authors acknowledge support from the
COST Action MP0903 ``Nanoalloys as Advanced Materials: From Structure to Properties and Applications''.

\end{document}